\documentclass[12pt]{iopart}
\usepackage{latexsym}
\usepackage{mathrsfs}
\usepackage{amsfonts}
\usepackage{epsfig}
\usepackage{setspace}
\usepackage{graphicx}
\begin{document}

\title{Phase Transitions and Dynamics in Bulk and Interfacial Water}

\author{G. Franzese,$^1$ A. Hernando-Mart\'{\i}nez,$^1$ P. Kumar,$^2$ M. G. Mazza,$^3$
  K. Stokely,$^3$  E. G. Strekalova,$^3$ F. de los Santos$^4$, H. E. Stanley,$^3$
}

\address{$^1$Departament de F\'{\i}sica Fonamental,
Universitat de Barcelona, Diagonal 647, Barcelona 08028, Spain\\
$^2$Center for Studies in Physics and Biology,
Rockefeller University, 1230 York Ave, New York, NY 10021 USA\\
$^3$Center for Polymer Studies and Department of Physics\\ Boston
  University, Boston, MA 02215 USA\\
$^4$Departamento de Electromagnetismo y F{\'\i}sica de la
Materia, Universidad de Granada, Fuentenueva s/n, 18071 Granada,
Spain}

\ead{gfranzese@ub.edu}

\begin{abstract}
New experiments for water at the surface of proteins at very low
temperature display intriguing dynamic behaviors.
The extreme conditions of these experiments make it difficult to explore the wide range of thermodynamic state points needed to offer a suitable interpretation.  Detailed simulations suffer the same problem, where equilibration times at low temperature become unreasonably long.
We show how Monte Carlo simulations and mean
field calculations of a tractable model of water help interpret the experimental results.
Here we summarize the results for bulk water and investigate
the thermodynamic and dynamic properties of supercooled water at
an interface.
\end{abstract}

\maketitle

\section{Introduction}

Water is essential in biology, because it participates in
nearly every process necessary for life (including cell metabolism,
transport of nutrients and residues, protein conformation changes,
etc.), and is the most common solvent in chemistry.
It regulates a large variety of processes, including
atmospheric phenomena,
the formation of geophysical structures,
the propagation of cracks in stones and cement,
the sliding of glaciers,
the transport in  plants,
and is ubiquitous in the universe as ice in the interstellar space.
In all these examples the properties of water are essential to
understand what is observed.

Nevertheless, water has proven to be a complex puzzle to many
researchers for its
anomalous thermodynamic and dynamic properties at room temperature and
pressure.
 For example, by decreasing temperature $T$ at  pressure
$P=1$~atm, water's volume fluctuations, proportional to the isothermal
compressibility $K_T$, increase below $T=46^o$C, and entropy
fluctuations, proportional to the isobaric specific heat $C_P$
increase below $T=35^o$C, while in normal liquid any fluctuation
decreases when $T$ is decreased
\cite{debenedetti_stanley,Angell1973science,SPEEDY1976}.
These water's anomalies grow upon cooling and increase in number. For
example below $T=4^o$C the cross-fluctuation of volume and entropy,
proportional to the isobaric thermal expansion coefficient $\alpha_P$,
becomes negative \cite{hare1986}, while it is always positive in normal liquids
where the entropy decreases when the volume decreases \cite{FS2007}.

By decreasing $T$ even more, it is experimentally possible to supercool
liquid water down to $T_H=-41^o$C at 1 atm and to $T_H=-92^o$C at 2000
atm, where the liquid is metastable with respect to crystal phases
 \cite{debenedetti_stanley}. These extreme conditions are not unusual
 in nature, where water exists in its liquid form at $-20^o$C in
 insects, $-37^o$C in clouds or $-47^o$C in plants \cite{debenedetti-book}.
Below $T_H$ the homogeneous nucleation
of  crystal ice occurs in a time too short to allow any
measurements. But even the crystal phase of water is not simple. In
fact, water is a polymorph with at least sixteen forms of crystal ices, the last
one, Ice XV, was discovered in 2009 \cite{IceXV}.

However, at very low $T$, crystal water is not the only possible kind
of ice. By rapidly quenching liquid water below $-123^o$C
\cite{Bruggeller1980}, or by condensing the vapour at low $T$
\cite{Jenniskens1994}, or by compressing crystal ice at low $T$
\cite{Mishima1985}, or by irradiation (with ions for example \cite{Strazzulla1992}),
it is possible to solidify water as an amorphous, or glass, i. e. a form
that has the elastic properties of a solid, but the structure of a
liquid with no long-range order \cite{Loerting_Giovambattista}.
As for the crystal state, the amorphous state of water is also not
unique. Water is a polyamorph with at least three different amorphous
states: low--density  amorphous (LDA), discovered in 1935 \cite{Burton1935},
high--density amorphous (HDA), discovered in 1984  \cite{Mishima1984},
and very high-density amorphous ice (VHDA), discovered in 2001
\cite{VHDA}.

All these anomalies of water are a consequence of the properties of the
hydrogen bond network that they form. The hydrogen bond interaction is
characterized by a preferred geometrical configuration,
that at low $T$ and $P$ is approximately a tetrahedron
of four molecules around a central one, with an angle varying around
$106.6^o$ (slightly smaller than a tetrahedral angle of $109.47^o$)
and a distance oscillating around $2.82$~\AA \cite{FS2007}.
The local arrangement, including the number of nearest neighbours,
can change with $T$ and $P$. In particular, in 2000
Soper and Ricci observed at $268$~K, compressing from
$26$ to $400$~MPa, a continuous transformation from
low--density liquid (LDL) local arrangement of water with an open,
hydrogen-bonded tetrahedral
structure, to high--density liquid (HDL) local arrangement with
nontetrahedral O-O-O angles and a
collapsed second coordination shell with broken hydrogen bonds, and  a
change in density of about 73\% \cite{Soper-Ricci-2000}.

\subsection{Thermodynamic Interpretations of Water Behavior}

All the above results are consistent with theories
that propose different mechanisms and different phase behaviors in the
supercooled region. They can be summarized in four possible scenarios for the
$P-T$ phase diagram.

(i) In the {\it stability limit} (SL) scenario~\cite{Speedy82}
the behavior of the superheated liquid spinodal, i. e. the limit of
stability of the liquid with respect to the gas, and the stretched
water, i. e. water under tension as in a plant fibers, are related. In
particular, it is hypothesized that the limits of stability of these
two regions are continuously connected  at negative pressure, forming a
re-entrant  curve toward the positive $P$ region below  $T_H(P)$.
The response functions, including $K_T$, $C_P$ and
$\alpha_P$, diverge when $T$ is decreased a positive $P$
as a consequence of the approaching of the re-entrant spinodal line.

(ii) In the \emph{liquid--liquid critical point} (LLCP) scenario
\cite{llcp} it is hypothesized the existence of a LDL--HDL
first--order phase transition line with negative slope in the $P-T$
plane and  terminating in a critical point $C'$.  Below the
critical pressure $P_{C'}$ the response functions increase on
approaching the Widom line (the locus of correlation length maxima
emanating from $C'$ into the one--phase region) \cite{FS2007}, and for $P>P_{C'}$ by
approaching the HDL--to--LDL spinodal line.  The possibility
with $P_{C'}<0$ have also been proposed~\cite{Tanaka96}.

(iii) In the \emph{singularity--free} (SF) scenario~\cite{Sastry1996} it is
hypothesized that the low-$T$ anticorrelation between volume and
entropy gives rise to response functions that increase upon cooling
and display maxima at non--zero $T$, but do not display any singular
behavior. Specifically, Sastry et al.~\cite{Sastry1996} show that this
is a direct consequence of the fact that water's line of
temperatures of maximum density (TMD) has a negative slope in
the $(T,P)$ plane.

(iv) In the~\emph{critical--point free} (CPF) scenario
\cite{Angell2008} it is
hypothesized that a first--order phase transition line separates two
liquid phases and extends to $P<0$ toward the superheated
limit of stability of liquid water.
This scenario effectively predicts a continuous locus of stability
limit spanning the superheated, stretched and supercooled state,
because the spinodal associated with the first--order transition will
intersect the liquid--gas spinodal at negative pressure. No critical
point is present in this scenario.

Since all these scenarios are consistent with the available
experimental data, a natural
question is if we can design an experiment that would
discriminate among them. Unfortunately, many scientists have
discovered that an answer to this question is a difficult challenge
\cite{Angell2008}.
In fact, experiments on bulk water are hampered by
freezing below $T_H$, and no measurements on bulk liquid water can
be performed with our present technology
below this temperature. Nevertheless, the different proposed theories
have different implications on phenomena such as the cold denaturation
(and stop of activity) of proteins at low $T$, an important issue in
cryopreservation, cryonics, cryostasis and cryobiology.

\subsection{Interfacial Water}

One possible strategy to probe supercooled water at very low $T$ is to
consider water at an interface.
Water adsorbed onto the
surface of proteins or confined in nanopores
freezes at much lower $T$ than bulk water, giving
access to a low-temperature region where interfacial water is still
liquid, while bulk water would not be \cite{MBBCMC2007}.
In many cases of interest for practical purposes in biology, geology
or industrial applications, water is hydrating a surface or is
confined. As a consequence, fundamental research in physics and
chemistry has been performed in recent years with  experiments
\cite{Angell2008,Granick2008,FLMZC2009,Ricci2009,Soper2008,Bellissent2008,
Vogel2008,PKS2008,MBBCMC2007,CLFBFM2006}, theories and simulations
\cite{Chandler2005,Dill2005,GLRD2008,GR2007}.

During the last years experiments on
water in Vycor micropores \cite{PhysRevE.51.4558},  in nanopores of
MCM-41 silica
\cite{Takahara:1999sc,faraone:3963,Sow-HsinChen08292006},  of sol--gel
silica glass
\cite{Crupi:2002ms}, of NaA zeolites \cite{Crupi:2004xi}, or of
double-wall carbon nanotubes \cite{Chu:2007hc}
have
contributed to the investigation of water dynamics in  confinement.
In particular, confinement in hydrophilic MCM-41 silica
nanopores of $1.8$ and $1.4$~nm diameter allows to study water dynamics down to
$200$~K where quasielastic neutron scattering reveals a crossover at
$T\approx 225$~K in
the average translational relaxation time from a non-Arrhenius
behavior at high $T$ to an Arrhenius behavior at low $T$
\cite{faraone2004}.
A similar crossover is also observed for the self-diffusion coefficient
of water by nuclear magnetic resonance at $T\approx 223$~K
\cite{mallamace2006}.
By increasing from $400$~bars to $1600$~bars
the external pressure applied on a sample of MCM-41 silica
nanopores with  $1.4$~nm diameter at full hydration level of $0.5$
g of H$_2$O per  g of silica, it has been observed that the
crossover occurs at lower $T$, reaching $T\approx 200$~K at $P=1600\pm
400$~bars and disappears at higher $P$ \cite{Liu2005}.

Quasielastic neutron scattering experiments show the same
crossover for the average translational relaxation time of
at least three different systems: 
(i) water hydrating lysozymes, at hydration level
$h=0.3$ g of H$_2$O per g of dry lysozyme, for $T\approx
220$~K, a temperature below which the protein has a glassy dynamics
\cite{CLFBFM2006},
(ii) DNA hydration water, at hydration level corresponding to about 15
water molecules per base pairs, for $T\approx
222$~K, at which DNA displays the onset of anharmonic molecular
motion \cite{chen-JCP06}, and 
(iii) RNA hydration water, at a similar hydration level, for $T\approx
220$~K, where both RNA and its hydration water exhibit a sharp change
in slope for  the mean-square displacements of the hydrogen atoms
\cite{chu:011908},

All these results can be interpreted as a consequence of a structural
rearrangement of water molecules associated with a LDL-HDL critical
point \cite{Liu2005}. In fact, along the Widom line in the
supercritical region of  the LDL-HDL critical point, the changes in
the hydrogen bond network are consistent with the dynamic behavior
observed in the experiments, as we will discuss in the following sections.

This interpretation has been criticized \cite{cerveny:189802} on the base of similar
crossover observed for water confined in molecular sieves
\cite{Jansson:2003hk} or for water mixtures \cite{sudo:7332} and water
solutions \cite{cerveny:194501}. It has also been proposed a possible
interpretation as a consequence of the dynamics of (Bjerrum-type) defects
due to orientationally disordered water molecules that are hydrogen
bonded to less than four other water molecules
\cite{swenson:189801,SwensonSitges}.

$^2$H-NMR studies on hydrated proteins, at a
comparable hydration level as in \cite{Liu2005},
show no evidence for the crossover at $220$~K and
indicate that  water performs thermally activated and distorted
tetrahedral jumps at $T<200$~K, which
may be related to a universal defect diffusion \cite{Vogel2008}.
Also, dielectric spectroscopy studies of hydrated protein show a
smooth temperature variations of conductivity at $220$~K and ascribe
the crossover observed in neutron scattering to a secondary
relaxation that splits from the main structural relaxation
\cite{PKS2008,Khodadadi:2008aq}.

On the other hand, numerical simulations for bulk water show that
crossing the Widom line emanating from a LDL-HDL critical point, the
structural change in water is maximum, as emphasized by the maximum in
specific heat, and the diffusion constant has a crossover \cite{Xu2005}.
This result is observed also in
water hydrating lysozyme or DNA, where the dynamic transition
of the macromolecules
occurs at the temperature of dynamic crossover in the
diffusivity of hydration water and also coincides with the maximum
change in water structure \cite{kumar:177802}.

A crossover from high-$T$ non-Arrhenius to low-$T$ Arrhenius behavior
is observed also in simulation of water hydrating lysozyme powder
in the translational correlation time and in
the inverse of the self-diffusion constant, in agreement with the
neutron scattering experiments, at about $223$~K \cite{Lagi:2008kr}.
The activation energy for the Arrhenius regime is found to be of about
$0.15$~eV, as in the neutron scattering experiments
\cite{Lagi:2008kr}. Also, simulations of water hydrating
elastin-like and collagen-like peptides show this crossover, but with
a weaker change in the slope and an Arrhenius activation energy
of about $0.43$~eV, consistent with dielectric spectroscopy and
nuclear magnetic resonance studies
\cite{PhysRevLett.93.245702,Vogel2008}.

It is therefore difficult not only in the experiments, but also in the
models to get a clear answer about the relevant dynamic mechanisms
and their relation with the thermodynamics in water at interfaces.
Moreover, the relation between confined water and bulk water remains
not fully understood. For this reason models that are tractable with a
theoretical approach are particularly appealing in this context.
With these models, in fact, simulations can be compared with analytic
calculations to develop a consistent theory.

\section{Cooperative Cell Model for a Monolayer of Water}

We consider the case of water in two dimensions (2D). This case can be
considered as an extreme confinement of one single layer of water
between two repulsive (hydrophobic) walls when the distance between
the walls is such to inhibit the crystal formation
\cite{KSBS2007}. In fact, it has been shown that the relevant
parameter to avoid the transition to a crystal phase is the distance
between the confining wall and not the characteristics of the
hydrophobic interaction with the wall \cite{KSBS2007}.

Another case in which the study of a monolayer of water is relevant is
when a substrate of protein powder is, on average, hydrated only by a single
layer of water, and the proteins do not undergo any configurational
transformation and/or large scale motion \cite{Mazza:2009}. In these conditions, for a
hydrophilic protein surface, we can assume that the effect of the water--protein
interaction is to attract water on a surface that, by
constraining the water molecule positions,
inhibits its crystallization.

A very desirable feature of a model for a liquid is transferability.
The parameters and effective interactions of a model are optimized to
precisely reproduce static and dynamic properties of the liquid at one
particular thermodynamic state point. The quality of the model is
measured by the range of validity of its predictions in other state
points. Unfortunately, there is no water model that is truly
transferable, nor can reproduce all the properties of water~\cite{Guillot02}.
Many routes have been explored to solve this issue.
Molecular polarizability~\cite{Lamoureux-LCP03,Mankoo-JCP08} is
one way to introduce effects not considered by standard pairwise
additive potentials. However, polarizable models are computationally
very expensive and provide only a partial solution~\cite{Piquemal-JPC07}.
An alternative way is to include many--body effects into the
potential. In the following we define a model with an
effective many--body interaction introduced through a cooperative
hydrogen bond term.

\subsection{Definition of the Model with Cooperative Interaction}

We consider $N$ molecules in a volume $V$ with periodic boundary
conditions (p.b.c.) in two dimensions, and the size of about one single molecule (and
no p.b.c.) in the third dimension.
We initially consider the case in which the molecules are distributed in a
homogeneous way, with each molecule $i\in[1,N]$ occupying the same
volume $V/N$ larger than a hard--core volume $v_0\approx 102$~\AA$^3$ due to
short-range electron clouds repulsion.
We consider the case in which each molecule has coordination number
four, consistent with the tendency of a water molecule to minimize its
energy by forming four hydrogen bonds.

The interaction Hamiltonian for water molecules
is~\cite{FS2002,FS-PhysA2002,FMS2003,FS2007}
\begin{equation}
  \mathscr{H} =
U_0(r)
-J \sum_{\langle i, j \rangle} \delta_{\sigma_{ij},\sigma_{ji}}
-J_\sigma
\sum_{(k,l)_i} \delta_{\sigma_{ik},\sigma_{il}}
\label{hamilt}
\end{equation}
where $U_0(r)$ denotes the sum of all the isotropic interactions
(e. g. van der Waals) between molecules at distance
$r\equiv(V/N)^{1/d}$, represented by a Lennard--Jones potential with
attractive energy $\epsilon$ plus a
hard--core at distance
$r_0\equiv(v_0)^{1/d}$.

The second term (with $\delta_{a,b}=1$ if $a=b$ and
$\delta_{a,b}=0$ otherwise, and $\langle i,j\rangle$ denoting that $i$ and $j$
are nearest--neighbors) accounts for the directional contribution to the
hydrogen bond energy with strength $J$, where
$\sigma_{ij}=1,\ldots,q$
is a (Potts) variable
representing the orientational state of the hydrogen (or the lone e$^-$) of
molecule $i$ facing the lone e$^-$ (or the hydrogen, respectively) of the
molecules $j$.
For the sake of simplicity we do not distinguish between
hydrogen and lone e$^-$, associating to each molecule four equivalent bond
indices $\sigma_{ij}$.
We choose the parameter $q$ by selecting 30$^o$ as the maximum
deviation from a linear bond, i. e. the O---H....O angle is less than
30$^o$, as estimated from Debye-Waller factors
\cite{Teixeira1990,Luzar-Chandler96}. Hence, $q\equiv 180^o/30^o=6$ and
every molecule has $q^4=6^4\equiv 1296$ possible orientations.
The effect of choosing a different value for $q$ has been analyzed
in~\cite{FS2007}.

The third term (with
$(k,l)_i$ indicating each of the six different pairs of the four bond indices
of molecule $i$) represents an interaction accounting for the hydrogen bond
cooperativity and giving rise to the O--O--O correlation~\cite{Ricci2009}, locally
driving the molecules toward an ordered (tetrahedral in the bulk)
configuration with lower energy.

By defining the energy per hydrogen bond
(between $\sigma_{ij}$ and $\sigma_{ji}$) as the sum of the interactions in
which two bonded molecules ($i$ and $j$) are participating, we obtain $E_{\rm
HB}=\epsilon+J+m J_\sigma/2$, where $m=0,\dots , 6$ is the number of
cooperative interactions in which that bond variables ($\sigma_{ij}$ and
$\sigma_{ji}$) are partaking.
If we choose as parameters
$\epsilon=5.8$~kJ/mol
(consistent with the value $5.5$ kJ/mol of
the estimate of the van der Waals attraction based on isoelectronic molecules
at optimal separation \cite{Henry2002}), $J=2.9$ kJ/mol and $J_\sigma=0.29$ kJ/mol,
the values of
$E_{\rm HB}$ ranges between $8.70$ and $9.6$ kJ/mol depending on $m$. However,
a definition of $E_{\rm HB}$ based on a cluster of $5$ or $8$ bonded molecules
in $d=$3-dimensions increases this range up to $17$ or $18$ kJ/mol, respectively.
Therefore, $E_{\rm HB}$ depends on the environment (the value of $m$ and the
number of molecules bonded in a cluster), as observed in computer simulation
of the crystalline phases of ice \cite{Baranyai05}, and has values within the
range 6.3 \cite{Smith2004}--- 23.3 kJ/mol \cite{Suresh00}, proposed on the
base of experiments.

Experiments show that formation of the hydrogen bonds leads to an open
---locally tetrahedral---
structure that induces an increase of volume per
molecule~\cite{De03,Soper-Ricci-2000}.
This effect is incorporated in the model by considering that a full
bonded molecule, i. e. a molecule with four hydrogen bonds, has a
molecular volume larger than a non-bonded molecule by an amount
\begin{equation}
\label{d_vol_mol}
\Delta v \equiv 4 v_{\rm HB},
\end{equation}
\noindent
where $v_{\rm HB}$ is the volume increase per H bond.
Hence, if
\begin{equation}
N_{\rm HB}\equiv
\sum_{\langle i,j \rangle} n_i n_j  \delta_{\sigma_{ij},\sigma_{ji}}
\end{equation}
\noindent
is the total number of hydrogen bonds in the system, the hydrogen bond
contribution to the total volume is
\begin{equation}
\label{d_vol}
\Delta V \equiv  N_{\rm HB} v_{\rm HB}.
\end{equation}
\noindent
We adopt
$r_0=2.9$~\AA~consistent with the expected value of
the van der Waals radius \cite{Narten67}, and
$v_{\rm HB}=0.5 v_0$, with $v_0\equiv r_0^3$, corresponding to a
 maximum hydrogen bond distance of about $3.3$~\AA,
consistent with
the range of a water molecule's first coordination shell, $3.5$~\AA, as
determined from the oxygen-oxygen radial distribution function
\cite{Soper1986}.

\section{The Phase Diagram}

The model is studied using both mean--field (MF) analysis and Monte
Carlo (MC) simulations. The MF approach has been describe in details
in Ref.s \cite{FS2007,Odessa2009}. It consists of expressing
the molar Gibbs free energy in terms of an exact partition function for
a portion of the system made of a treatable number of degrees of
freedom. We take into account the effect of all the rest of the system as a
mean field acting on the border of this portion
\cite{FS2007,FS2002,FS-PhysA2002,FMS2003,Odessa2009,FYS2000,Franzese:2005sf,KumarFS2008,KFS2008,KumarFBS2008,FranzeseSCKMCS2008}.

MC simulations are performed at constant $N$, $P$, $T$, allowing the
volume $V_{\rm MC}$ of the system to fluctuate as a stochastic
variable. The average distance $r$ between the molecules is then
calculated as $r/r_0\equiv (V_{\rm MC}/v_0N)^{1/d}$, where $d=2$ in two
dimensions. The total volume of the system is by definition
\begin{equation}
\label{vol}
V \equiv  V_{\rm MC}+\Delta V,
\end{equation}
\noindent
where $\Delta V$ is the hydrogen bond volume contribution in Eq.(\ref{d_vol}).
Note that $\Delta V$ is not included in the calculation of $r$ to
avoid MF--type long--range correlation in volume fluctuations in the MC
simulations.

\begin{figure}
(a)
\includegraphics[width=7cm]{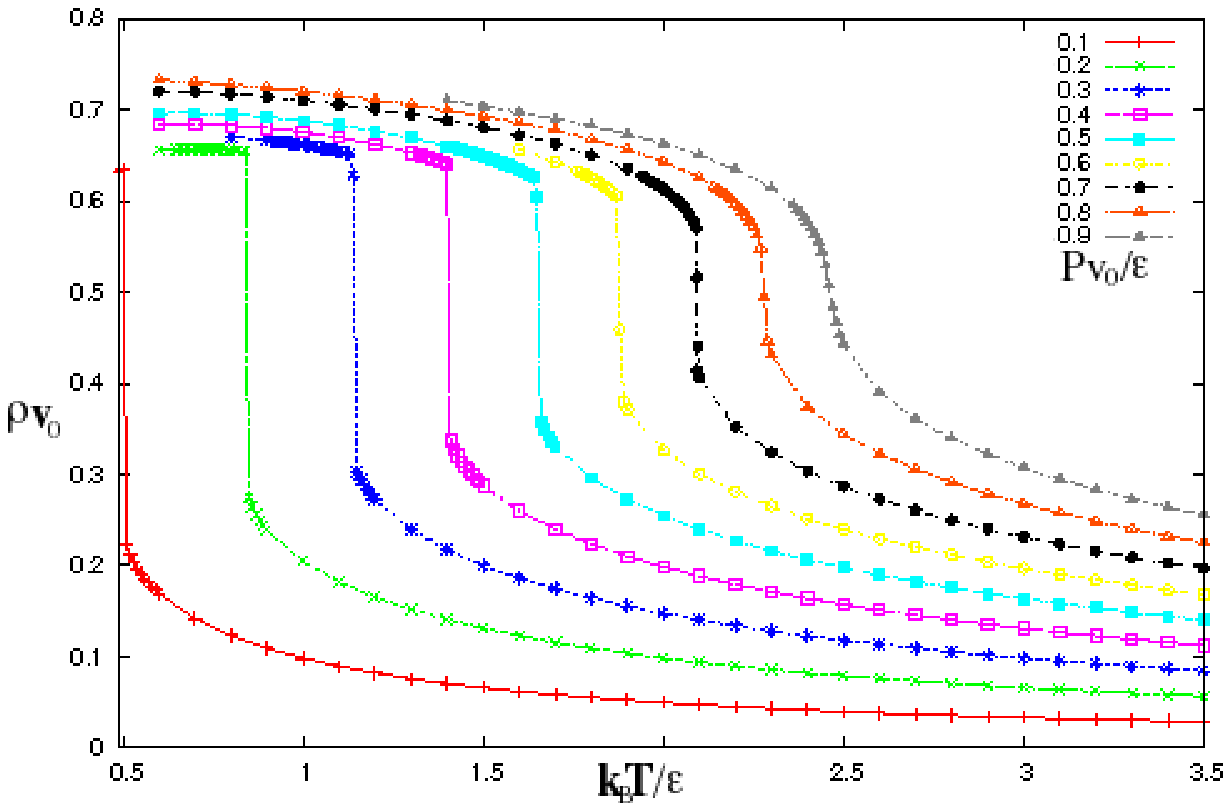}
(b)
\includegraphics[width=7cm]{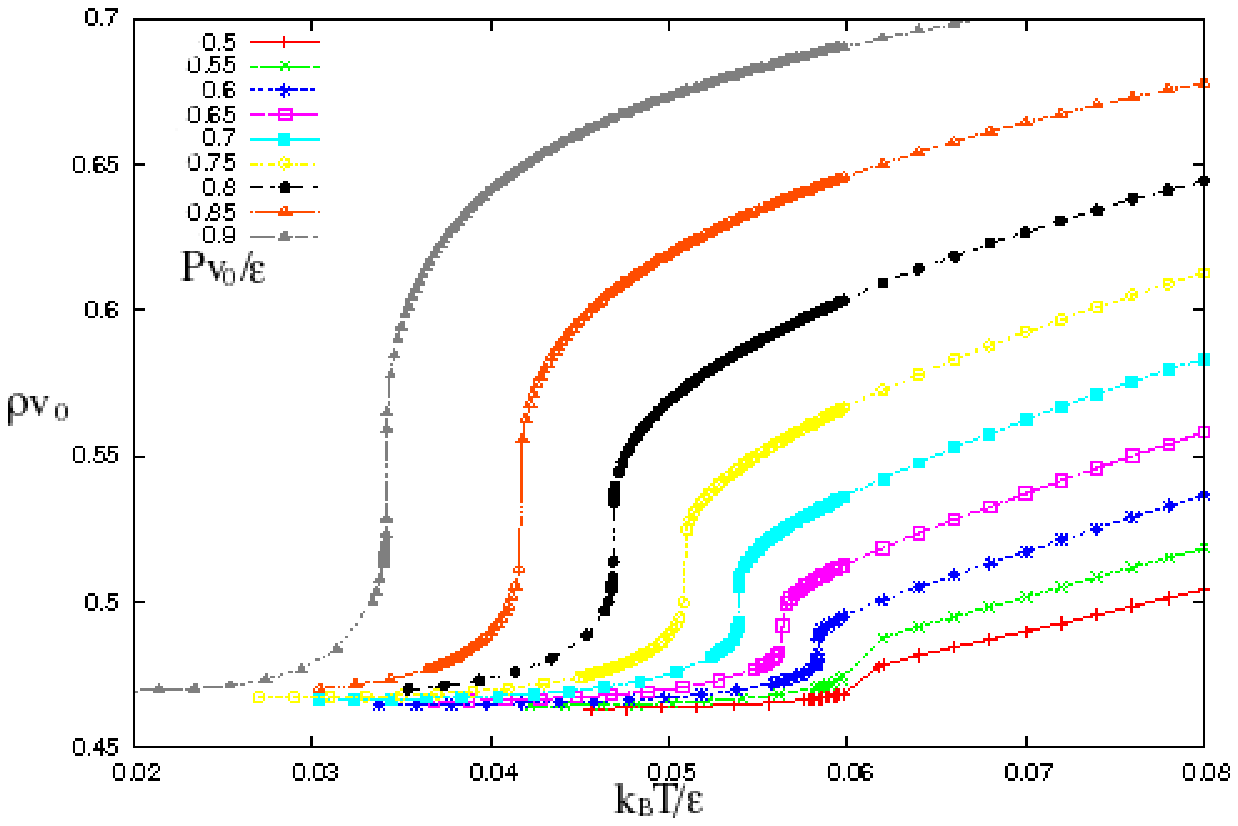}
\caption{The density $\rho$ (in units of $1/v_0$) as a function of the
  temperature $T$ (in units of $\epsilon/k_B$ where $k_B$ is the
  Boltzmann constant) for different values of  pressure $P$ (in units
  of $\epsilon/v_0$) as calculated from MC simulation of a system with
  $N=15625$ water molecules. The parameters of the model are
  $J/\epsilon=0.5$, $J_\sigma/\epsilon=0.05$ and $v_{\rm HB}/v_0=0.5$.
(a) At high $T$ and for (from bottom to top) values of $Pv_0/\epsilon$ from
  $0.1$ to $0.9$, we observe for $Pv_0/\epsilon<0.8$ a  discontinuity
in the density corresponding to the first-order gas-liquid phase
transition, with a critical $P$ at about $(0.75\pm
 0.05)\epsilon/v_0$ and critical $T$ at about $(2.2\pm  0.1)\epsilon/k_B$.
Note that if we choose as
model parameters $\epsilon=2.5$~kJ/mol and $r_0=3.2$~\AA,  we
get an estimate $P_{C'}=22.7\pm1.5$~MPa and $T_{C'}= 661\pm30$~K
consistent with the real water
critical point at about $22.064$~MPa  and $647$~K.
(b) At low $T$ and for (from bottom to top) $Pv_0/\epsilon$ from
 $0.5$ to $0.9$, for $Pv_0/\epsilon>0.55$ a discontinuity
in $\rho$ marks the first-order LDL-HDL phase transition.}
\label{fig1}
\end{figure}

For the parameters choice
$J/\epsilon=0.5$, $J_\sigma/\epsilon=0.05$ and $v_{\rm HB}/v_0=0.5$,
we find that the density $\rho$ as a function of $T$ at constant $P$
displays a discontinuous change at high $T$ and low $P$ corresponding
to the gas-liquid first order phase transition anding in a critical
point where the discontinuity disappears (Fig.~\ref{fig1}a). By
decreasing $T$, the density reaches a maximum, that in real water at
atmospheric pressure occurs at $4^o$C. At lower $T$, in the
supercooled state, and higher $P$ we
find another discontinuity in density, this time with a lower density
at lower $T$ (Fig.~\ref{fig1}b). The system at these supercooled $T$ displays a first
order phase transition from HDL to LDL, as hypothesized in the LLCP
scenario (Fig.~2).

%

\begin{figure}
\label{fig2}
\begin{center}
\includegraphics[width=8cm]{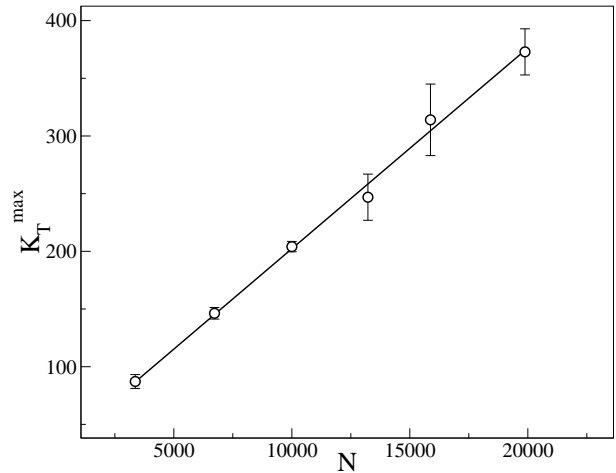}
\caption{The finite size behavior of the maximum of compressibility
  $K_T^{\rm max}$ as a function of the number of water molecules $N$
  for pressure $P=0.8\epsilon/v_0$ at low $T$
  shows a linear increase as expected at a first order phase
  transition.}
\end{center}
\end{figure}

\subsection{Effect of Hydrogen Bond Cooperativity on the Behavior of
  Water}

The experiments for confined water have boosted the debate over the
supercooled phase diagram of
water, motivating the proposal of the
CPF scenario hypothesized
by C. A. Angell \cite{Angell2008},
as described above.
This new scenario leads to questions such as
\begin{itemize}
 \item[(i)] How to
understand the new Angell hypothesis?
\item[(ii)] How to connect it to the other three
existing hypotheses?
\end{itemize}

A recent work by Stokely et al. \cite{Stokely2010}
succeeds in answering both questions (i) and (ii).
Specifically, it is shown that all four existing hypotheses are cases of
the cooperative water
model.  Thus no matter which hypothesis may be correct (if any
is correct), it is possible that the underlying mechanism is basically
the same---the thermodynamic properties of water can be accounted for by
considering two main contributions to the hydrogen bond interaction: (a)
the directional (or covalent) contribution (parametrized by $J$ in the
model) and (b) the three-body (or cooperative) contribution
(parametrized by $J_\sigma$ in the model).

By  MF calculations and MC simulations, Stokely et
al. \cite{Stokely2010} demonstrate
that the balance between contributions (a) and (b) determines which of
the four hypotheses presented in section 1.1. best describes
experimental facts. Since the
characteristic energy associated with these two contributions can be
estimated, the work allows to begin to validate or contradict each
hypothesis on an experimental basis.

\begin{figure}
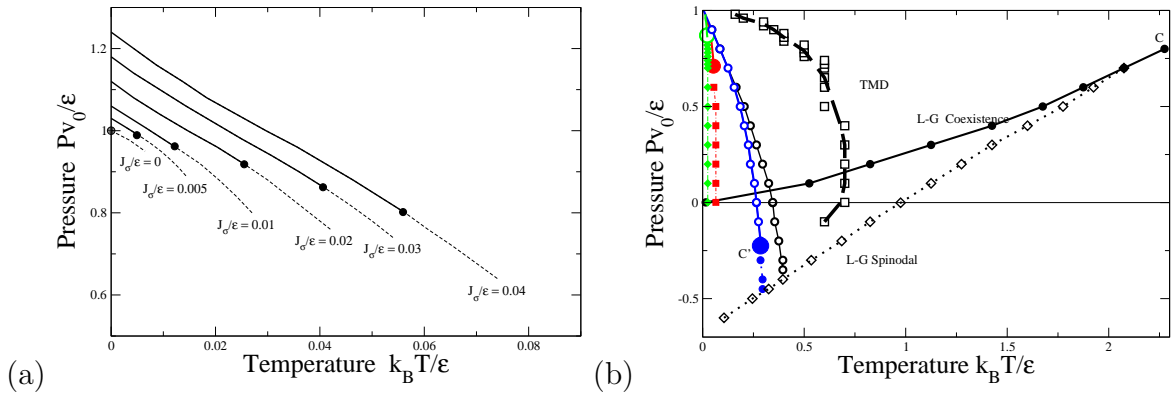

(a)
\includegraphics[width=7cm]{FIG-LIMIT.eps}
(b)
\includegraphics[width=7cm]{1-new6-PT-N10000.eps}
\caption{The pressure $P$ vs temperature $T$ phase diagram of the
  cooperative water model for different values of
  $J_\sigma/\epsilon$. (a) Mean field results showing the
low $T$ phase diagram with the LDL-HDL phase transition lines (solid
lines, where $K_T$ is discontinuous)  and Widom lines (dashed
lines, where $K_T$ has a finite maximum)
for varying $J_\sigma/\epsilon$ from (rightmost) 
0.04, 0.03, 0.02, 0.01, 0.005 and 0 (leftmost). For each value of
$J_\sigma/\epsilon$ solid circles indicate the
LDL-HDL  critical point  $C'$
where the response function (such as $K_T$) diverge. Hashed
circle indicates the state point at $T=0$ where $K_T$ diverges when
$J_\sigma/\epsilon=0$.
(b) The MC phase diagram for varying $J_\sigma/\epsilon$ for $N=10^4$
water molecules.
At high $T$ the system displays a liquid--gas first--order phase
transition line (continuous line with full circles) ending in a liquid--gas
critical point $C$ (full circle), from which departs the liquid-to-gas spinodal
line (dashed line with open diamonds). At lower $T$, the retracing line
with open squares marks the temperatures of maximum density (TMD) along
isobars. All these loci do not change in an appreciable way with the
value of $J_\sigma/\epsilon$. The phase diagram at lower $T$, instead,
show a strong dependence from $J_\sigma/\epsilon$. For
$J_\sigma/\epsilon=0.5$, we find for any $P$ above the spinodal line a
first-order phase transition line between a HDL (at high $P$) and a
LDL (at low $P$) phase (continuous line with open circles). This is the
CPF scenario \cite{Angell2008}. The analysis of the HDL-to-LDL (not shown) reveals that
the limit of stability of the HDL phase retraces to positive $P$ as in
the SF scenario \cite{Speedy82}. For $J_\sigma/\epsilon=0.3$ we observe that the
HDL-LDL phase transition ends to a critical point $C'$
(continuous line with open circles ending in a full large circle)
at negative $P$, as in the LLCP scenario suggested in Ref.
\cite{Tanaka96}. From $C'$ a Widom line (dashed line with full
circles) departs.
For $J_\sigma/\epsilon=0.05$, $C'$ occurs at positive $P$, has
hypothesized in Ref. \cite{llcp} and the Widom line (dashed line with
full squares) extends to lower $P$.
For $J_\sigma/\epsilon=0.02$, $C'$ approaches the $T=0$ axis, as well
as the Widom line  (dashed line with full diamonds), going toward the
limit of the SF scenario \cite{Sastry1996}.}
\label{fig3}
\end{figure}

Specifically, by fixing the parameters $J/\epsilon=0.5$ and $v_{\rm
  HB}/v_0=0.5$, and varying the parameter $J_\sigma/\epsilon$, it is
possible to observe that the cooperative model reproduces all four
scenarios of section 1.1.
The overall picture that emerges is one in which the amount of
cooperativity among H bonds ($J_\sigma/\epsilon$), in relation to the H bond
directional strength ($J/\epsilon$), governs the location of a LLCP, hence which
scenario is realized.
For zero cooperativity, the temperature $T_{C'}$ where $K_T^{\rm max}$
and $\alpha_P^{\rm max}$ diverge is at zero
temperature, and no liquid-liquid transition exists for
$T>0$--the SF scenario.
For very large cooperativity, $C'$ lies outside the region of stable liquid states,
and a liquid-liquid transition extends to the entire (supercooled
and superheated) liquid phase--the CPF/SL scenario.
For intermediate values of H bond cooperativity, $T_{C'}$ varies in a
smooth way between these two extremes--the LLCP scenario.
Due to the anticorrelation between the volume and entropy associated
to the H bonds, the larger $T_{C'}$, the smaller $P_{C'}$,
eventually with $P_{C'}<0$ for larger cooperativity. These cases are
summarized in Fig.~\ref{fig3}.

\section{Water at Interfaces}

To elucidate the relation between the protein dynamic crossover at
about $220$~K and the dynamic crossover observed for the average
translational correlation time in the first layer of protein
hydration water \cite{CLFBFM2006,chen-JCP06,chu:011908,Liu2005},
we perform MF calculations and MC simulations of the cooperative
model of water of section 2.1.
Since we are interested in cases at low humidity, we consider the case
of a water monolayer hydrating an immobile surface of
globular protein that, forcing the water molecules out of place with
respect to crystal configurations, inhibits the crystallization.
We focus on the hydrogen bonds dynamics, regardless if the hydrogen
bonds are formed  with the protein or among water molecules.

\subsection{The Hydrogen Bond Dynamics for Hydrated Proteins}

Following the work of Kumar et
al. \cite{KFS2008,KumarFS2008,KumarFBS2008}, Mazza et
al. \cite{Mazza:2009} study the orientational correlation time $\tau$ associated with the
hydrogen bond dynamics of the model is in section 2.1. They confirm the
occurrence of a dynamic crossover at a temperature of about
$k_BT/\epsilon\approx0.32$ corresponding to the $T$ of maximum
variation of the number of hydrogen bonds, that in turn corresponds to
the Widom line. They also confirm that the
crossover is from a non--Arrhenius behavior at high $T$ to another
non--Arrhenius behavior \cite{FranzeseSCKMCS2008}, that closely
resembles an Arrhenius behavior around the crossover.

These results are consistent with those from simulations of
other models for hydrated proteins where a crossover in the
translational dynamics is observed \cite{kumar:177802}.
The difference here is that (i) the crossover is for
the dynamics of the hydrogen bonds, (ii) in the cooperative
water model the crossover can be calculated from MF and an exact
relation can be found between the crossover and the Widom line, and
(iii) the model can test different
hypotheses. In particular, Kumar et al. have shown that the crossover
at $k_BT/\epsilon\approx0.32$ is
independent of whether water at very low temperature is characterized
by a LLCP or is SF. In fact, the
crossover is a consequence of the sharp
change in the average number of hydrogen bonds at the temperature of
the specific heat maximum, that occurs in both scenarios.
Kumar et al. were able also to make predictions about the
$P$--dependence of quantities characterizing the crossover at
$k_BT/\epsilon\approx0.32$: (i) the time scale of the crossover,
showing that it is independent of $P$ (isochronic crossover); (ii) the activation energy of
the apparent Arrhenius behavior at low $T$ and (iii) the crossover
temperature, showing that both (ii) and (iii) decrease linearly upon
increasing $P$ \cite{KFS2008}. These predictions have been confirmed by Chu et
al. \cite{Chu:2009hl,FranzeseSCKMCS2008} in a study on
the dynamics of a hydrated protein under moderately high pressures at
low temperatures using the quasielastic neutron scattering
method. They relate these predictions (i)--(iii) to the mechanical
response of the protein to an external force, that is the average elastic
constant calculated from the mean square displacement of the protein
atoms. In particular, the degree of ``softness'' of the protein,
related to the enzymatic activity, is preserved at lower $T$ if the
pressure is increased \cite{Chu:2009hl}. However,
 a criterion proposed in Ref. \cite{KFS2008} for discriminating
which scenario better describe water on the basis of the crossover at
$k_BT/\epsilon\approx0.32$, cannot be tested in the experiments since
the predicted difference between the two scenarios (of the order of
1\%) is within the error bars of the measurements
\cite{FranzeseSCKMCS2008}.

The answer to the puzzle of which of the scenarios better describe
water might be related to the very recent experimental discovery of
another crossover for the hydrogen bond $\tau$ to an Arrhenius
behavior at very low $T$, of the order of $180$~K at hydration $h=0.3$
g H$_2$O/g. This crossover has been observed by Mazza et al. at
$k_BT/\epsilon\approx0.07$ \cite{Mazza:2009}, in relation to an
ordering process of the hydrogen bonds leading to the HDL-LDL critical
point. The study has been possible thanks to the use of a highly
efficient cluster MC dynamics  \cite{MSSSF2009,Cataudella:1996ko}.
This very--low $T$ crossover would reduce even more the $T$ at
which the proteins preserve their ``softness'', essential for their
correct functionality.

\subsection{Water Monolayer in Hydrophobic Confinement}

By considering partially hydrated hydrophobic plates at a distance such
to inhibit the crystallization of water at low $T$, Franzese and de
los Santos \cite{FDLS2009} have show that water has a glassy behavior
\cite{Kumar:2006zi}
for both the translational and rotational degrees of
freedom when cooled down to a
low $P$. This result is consistent with simulations of TIP4P water
forming a quasi-2d amorphous when
confined in a hydrophobic slit pore with wall-to-wall separation  just
enough to accommodate two molecular layers  \cite{Koga00}.

At higher $P$ the hydrogen bond network builds up in a less
gradual way, allowing the system to equilibrate the rotational degrees
of freedom also at very low
$T$, but not the translational degrees of freedom. This effect is
emphasized by the appearance of many dehydrated regions
\cite{FDLS2009}, as also observed in water confined between two
protein-like hydrophobic flattened surfaces at distances ranging from
$0.4$ to $1.6$~nm \cite{GLRD2008}.

When $P$ is close to the LLCP value, the
cooperativity of the hydrogen bond network induces a strong
non-exponential behavior \cite{FC1999}
for the hydrogen bond correlation
function. However, both rotational and translational degrees of
freedom equilibrate within the simulation time. At higher $P$ the
rotational correlation function recovers the exponential behavior and
the diffusion of the system allows the formation a large dry cavity,
while the rest of the surface is well hydrated. It is interesting to
observe that the cooperative model allows to calculate, in the MF
approximation, the diffusion constant at any $T$ and $P$
\cite{dlSF2009}.

The hydrophobic confinement has effects also on water
thermodynamics.  It shifts the HDL-LDL phase transition to lower
temperature and lower pressure compared to bulk water when the
confinement is between plates \cite{truskett:2401,KBSGS2005}.
Moreover, it
shifts both the line of maximum density and  the liquid-to-gas
spinodal toward higher pressures and lower temperatures with respect
to bulk when the confinement is in a hydrophobic disordered matrix of
soft spheres \cite{GR2007}. This result is confirmed also in the
analysis performed by using the cooperative water model in confinement
between hydrophobic hard spheres \cite{Elena2010}. However, the effect
of the matrix on the HDL-LDL critical point is less clear and is
presently under investigation.

\section{Conclusions}

The effect of confinement is of great interest to biology, chemistry, and engineering, yet
the recent experimental and simulations results are object of an intense debate.
A better understanding of the physico--chemical properties of liquid
water at interfaces is important  to provide accurate predictions of the behavior of
biological molecules \cite{Ball08}, including the folding-unfolding transitions
seen in proteins \cite{PhysRevE.62.8449,Levy,Dias:2008sp}, and the
dynamical behavior of  DNA~\cite{chen-JCP06}. However,
it is still unclear whether such behaviors are inherent in the structure of such
molecules, or an effect of water in which they are always found, or due to
the interactions between the two.

To get insight into this subject the formulation of a model that
allows the development of a theory could be useful to find functional
relations connecting different observables. The advantage of this
approach is to have two independent ways of approaching the problem,
one theoretical and the other numerical.

We have presented here several recent results obtained with a
cooperative water model suitable for studies with mean field theories
and with $N$, $P$, $T$ simulations with thousands of molecules. The
model has been studies in the context of water monolayers on hydrated
proteins, between hydrophobic surfaces or in a hydrophobic matrix.

Some of the conclusions reached with this model are the following.

\begin{itemize}
\item
The different scenarios proposed to interpret the low-$T$ behavior of
water are instances of the same mechanism, with different values of
the directional (covalent) strength and the cooperative (many-body)
interaction of the  hydrogen bonds. The parameters that can be
estimated from the experiments suggest that the scenario with the LLCP
is the most plausible for water.

\item
Previous experiments showing one dynamic crossover in the water
monolayer hydrating proteins, RNA and DNA
are consistent with (at least) two scenarios.

\item
The possibility of a second dynamic crossovers detectable at lower $T$
and lower hydration level would be  consistent only with the LLCP
scenario, because its origin would be related to the ordering of the
hydrogen bond network.

\item
A consequence of the occurrence of a LLCP should be detectable when the
translational and rotational dynamics of water are studied for a
monolayer in a hydrophobic confinement. In particular, the rotational
dynamics should appear with the strongest non-exponential behavior in
the vicinity of the LLCP, as an effect of the cooperativity. Moreover,
the slow increase of the number of hydrogen bonds at ow $T$ and low
$P$ is the cause of the formation of an amorphous glassy state when
the confinement is such that to inhibit the crystallization of water.
Under this conditions, the dehydration of hydrophobic surfaces is
characterized by the appearance of heterogeneities  and cavitation.

\item
 The hydrophobic confinement affects the thermodynamics of water by
 lowering the $T$ and increasing the $P$ of the liquid-gas phase
 transition and of the TMD line. It also affects the LDL-HDL phase
 transition in a way that is possibly more complex.

\end{itemize}

All these results are potentially relevant in problems such as the
protein denaturation or the protein aggregation. Works are in
progress to underpin and build up a theory of water at interfaces that
could help us to acquire a better understanding of these subjects.

\section*{Acknowledgments}
This work was partially supported by the Spanish Ministerio de Ciencia
e Innovaci\'on
Grants Nos. FIS2005-00791 and FIS2009-10210 (co-financed FEDER), Junta
de Andaluc\'ia Contract No. FQM 357.

\section*{References}

\bibliographystyle{./iopart-num}

\begin{thebibliography}{10}
\expandafter\ifx\csname url\endcsname\relax
  \def\url#1{{\tt #1}}\fi
\expandafter\ifx\csname urlprefix\endcsname\relax\def\urlprefix{URL }\fi
\providecommand{\eprint}[2][]{\url{#2}}

\bibitem{debenedetti_stanley}
Debenedetti P~G and Stanley H~E 2003 {\em Physics Today\/} {\bf 56} 40--46

\bibitem{Angell1973science}
Angell C~A and Tucker J~C 1973 {\em Science\/} {\bf 181} 342--344

\bibitem{SPEEDY1976}
Speedy R~J R~J and Angell C~A {1976} {\em {J. Chem. Phys.}\/} {\bf {65}}
  {851--858}

\bibitem{hare1986}
Hare D~E and Sorensen C~M 1986 {\em The Journal of Chemical Physics\/} {\bf 84}
  5085--5089

\bibitem{FS2007}
Franzese G and Stanley H~E 2007 {\em Journal of Physics-Condensed Matter\/}
  {\bf 19} 205126

\bibitem{debenedetti-book}
Debenedetti P~G 1996 {\em Metastable Liquids. Concepts and Principles\/}
  (Princeton, NJ: Princeton University Press)

\bibitem{IceXV}
Salzmann C~G, Radaelli P~G, Mayer E and Finney J~L 2009 {\em Physical Review
  Letters\/} {\bf 103} 105701

\bibitem{Bruggeller1980}
Bruggeller P and Mayer E 1980 {\em Nature\/} {\bf 288} 569--571

\bibitem{Jenniskens1994}
Jenniskens P and Blake D 1994 {\em Science\/} {\bf 265} 753--756

\bibitem{Mishima1985}
Mishima O, Calvert L and Whalley E {1985} {\em {Nature}\/} {\bf {314}} {76--78}

\bibitem{Strazzulla1992}
Strazzulla G, Baratta G~A, Leto G and Foti G 1992 {\em EPL (Europhysics
  Letters)\/} {\bf 18}

\bibitem{Loerting_Giovambattista}
Loerting T and Giovambattista N 2006 {\em J. Phys: Cond. Mat.\/} {\bf 18}
  R919--R977

\bibitem{Burton1935}
Burton E~F and Oliver W~F 1935 {\em Proc. R. Soc. Lond. A\/} {\bf 153} 166--172

\bibitem{Mishima1984}
Mishima O, Calvert L and Whalley E 1984 {\em Nature\/} {\bf 310} 393--395

\bibitem{VHDA}
Loerting T, Salzmann C, Kohl I, Mayer E and Hallbrucker A {2001} {\em {Physical
  Chemistry Chemical Physics}\/} {\bf {3}} {5355--5357}

\bibitem{Soper-Ricci-2000}
Soper A and Ricci M {2000} {\em {Physical Review Letters}\/} {\bf {84}}
  {2881--2884}

\bibitem{Speedy82}
Speedy R~J 1982 {\em The Journal of Physical Chemistry\/} {\bf 86} 3002--3005

\bibitem{llcp}
Poole P, Sciortino F, Essmann U and Stanley H {1992} {\em {Nature}\/} {\bf
  {360}} {324--328}

\bibitem{Tanaka96}
Tanaka H 1996 {\em Nature\/} {\bf 380} 328

\bibitem{Sastry1996}
Sastry S, Debenedetti P~G, Sciortino F and Stanley H~E 1996 {\em Physical
  Review E\/} {\bf 53} 6144--6154

\bibitem{Angell2008}
Angell C~A 2008 {\em Science\/} {\bf 319} 582--587

\bibitem{MBBCMC2007}
Mallamace F, Branca C, Broccio M, Corsaro C, Mou C~Y and Chen S~H 2007 {\em
  Proceedings of the National Academy of Sciences\/} {\bf 104} 18387--18391

\bibitem{Granick2008}
Granick S and Bae S~C 2008 {\em Science\/} {\bf 322} 1477--1478

\bibitem{FLMZC2009}
Faraone A, Liu K~H, Mou C~Y, Zhang Y and Chen S~H {2009} {\em {Journal Of
  Chemical Physics}\/} {\bf {130}}

\bibitem{Ricci2009}
Ricci M~A, Bruni F and Giuliani A 2009 {\em Faraday Discussussion\/} {\bf 141}
  347--358

\bibitem{Soper2008}
Soper A~K 2008 {\em Molecular Physics\/} {\bf 106} 2053 -- 2076

\bibitem{Bellissent2008}
Bellissent-Funel M~C 2008 {\em Journal of Physics: Condensed Matter\/} {\bf 20}
  244120

\bibitem{Vogel2008}
Vogel M {2008} {\em {Physical Review Letters}\/} {\bf {101}} 225701

\bibitem{PKS2008}
Pawlus S, Khodadadi S and Sokolov A~P 2008 {\em Physical Review Letters\/} {\bf
  100} 108103

\bibitem{CLFBFM2006}
Chen S~H, Liu L, Fratini E, Baglioni P, Faraone A and Mamontov E {2006} {\em
  {Proceedings of the National Academy of Sciences of the United States of
  America}\/} {\bf {103}} {9012--9016}

\bibitem{Chandler2005}
Chandler D 2005 {\em Nature\/} {\bf 437} 640--7

\bibitem{Dill2005}
Dill K~A, Truskett T~M, Vlachy V and Hribar-Lee B 2005 {\em Annual Review of
  Biophysics and Biomolecular Structure\/} {\bf 34} 173--199

\bibitem{GLRD2008}
Giovambattista N, Lopez C~F, Rossky P~J and Debenedetti P~G {2008} {\em
  {Proceedings of the National Academy of Sciences of the United States of
  America}\/} {\bf {105}} {2274--2279}

\bibitem{GR2007}
Gallo P and Rovere M 2007 {\em Physical Review E (Statistical, Nonlinear, and
  Soft Matter Physics)\/} {\bf 76} 061202--7

\bibitem{PhysRevE.51.4558}
Bellissent-Funel M~C, Chen S~H and Zanotti J~M 1995 {\em Phys. Rev. E\/} {\bf
  51} 4558--4569

\bibitem{Takahara:1999sc}
Takahara S, Nakano M, Kittaka S, Kuroda Y, Mori T, Hamano H and Yamaguchi T
  1999 {\em The Journal of Physical Chemistry B\/} {\bf 103} 5814--5819

\bibitem{faraone:3963}
Faraone A, Liu L, Mou C~Y, Shih P~C, Copley J~R~D and Chen S~H 2003 {\em The
  Journal of Chemical Physics\/} {\bf 119} 3963--3971

\bibitem{Sow-HsinChen08292006}
Chen S~H, Mallamace F, Mou C~Y, Broccio M, Corsaro C, Faraone A and Liu L 2006
  {\em Proceedings of the National Academy of Sciences\/} {\bf 103}
  12974--12978

\bibitem{Crupi:2002ms}
Crupi V, Majolino D, Migliardo P and Venuti V 2002 {\em The Journal of Physical
  Chemistry B\/} {\bf 106} 10884--10894

\bibitem{Crupi:2004xi}
Crupi V, Majolino D, Migliardo P, Venuti V, Wanderlingh U, Mizota T and Telling
  M 2004 {\em The Journal of Physical Chemistry B\/} {\bf 108} 4314--4323

\bibitem{Chu:2007hc}
Chu X~Q, Kolesnikov A~I, Moravsky A~P, Garcia-Sakai V and Chen S~H 2007 {\em
  Physical Review E (Statistical, Nonlinear, and Soft Matter Physics)\/} {\bf
  76} 021505--6

\bibitem{faraone2004}
Faraone A, Liu L, Mou C~Y, Yen C~W and Chen S~H 2004 {\em The Journal of
  Chemical Physics\/} {\bf 121} 10843--10846

\bibitem{mallamace2006}
Mallamace F, Broccio M, Corsaro C, Faraone A, Wanderlingh U, Liu L, Mou C~Y and
  Chen S~H 2006 {\em The Journal of Chemical Physics\/} {\bf 124} 161102

\bibitem{Liu2005}
Liu L, Chen S~H, Faraone A, Yen C~W and Mou C~Y 2005 {\em Phys. Rev. Lett.\/}
  {\bf 95} 117802

\bibitem{chen-JCP06}
Chen S~H, Liu L, Chu X, Zhang Y, Fratini E, Baglioni P, Faraone A and Mamontov
  E 2006 {\em The Journal of Chemical Physics\/} {\bf 125} 171103

\bibitem{chu:011908}
qiang Chu X, Fratini E, Baglioni P, Faraone A and Chen S~H 2008 {\em Physical
  Review E (Statistical, Nonlinear, and Soft Matter Physics)\/} {\bf 77} 011908

\bibitem{cerveny:189802}
Cerveny S, Colmenero J and Alegr\'{\i}a A 2006 {\em Physical Review Letters\/}
  {\bf 97} 189802

\bibitem{Jansson:2003hk}
Jansson H and Swenson J 2003 {\em The European Physical Journal E: Soft Matter
  and Biological Physics\/} {\bf 12} 51--54

\bibitem{sudo:7332}
Sudo S, Tsubotani S, Shimomura M, Shinyashiki N and Yagihara S 2004 {\em The
  Journal of Chemical Physics\/} {\bf 121} 7332--7340

\bibitem{cerveny:194501}
Cerveny S, Schwartz G~A, Alegr\'{\i}a A, Bergman R and Swenson J 2006 {\em The
  Journal of Chemical Physics\/} {\bf 124} 194501

\bibitem{swenson:189801}
Swenson J 2006 {\em Physical Review Letters\/} {\bf 97} 189801

\bibitem{SwensonSitges}
Swenson J, Jansson H and Bergman R {2008} {\em {Aspects Of Physical Biology:
  Biological Water, Protein Solutions, Transport And Replication}\/} ({\em
  {Lecture Notes In Physics}\/} vol {752}) ed {Franzese, G and Rubi, M} pp
  {23--42}

\bibitem{Khodadadi:2008aq}
Khodadadi S, Pawlus S, Roh J~H, Sakai V~G, Mamontov E and Sokolov A~P 2008 {\em
  Journal Of Chemical Physics\/} {\bf 128} 195106

\bibitem{Xu2005}
Xu L, Kumar P, Buldyrev S~V, Chen S~H, Poole P~H, Sciortino F and Stanley H~E
  2005 {\em Proceedings of the National Academy of Sciences of the United
  States of America\/} {\bf 102} 16558--16562

\bibitem{kumar:177802}
Kumar P, Yan Z, Xu L, Mazza M~G, Buldyrev S~V, Chen S~H, Sastry S and Stanley
  H~E 2006 {\em Physical Review Letters\/} {\bf 97} 177802

\bibitem{Lagi:2008kr}
Lagi M, Chu X, Kim C, Mallamace F, Baglioni P and Chen S~H 2008 {\em J Phys
  Chem B\/} {\bf 112} 1571--5

\bibitem{PhysRevLett.93.245702}
Cerveny S, Schwartz G~A, Bergman R and Swenson J 2004 {\em Phys. Rev. Lett.\/}
  {\bf 93} 245702

\bibitem{KSBS2007}
Kumar P, Starr F~W, Buldyrev S~V and Stanley H~E 2007 {\em Physical Review E
  (Statistical, Nonlinear, and Soft Matter Physics)\/} {\bf 75} 011202

\bibitem{Mazza:2009}
Mazza M~G, Stokely K, Pagnotta S~E, Bruni F, Stanley H~E and Franzese G Two
  dynamic crossovers in protein hydration water and their thermodynamic
  interpretation

\bibitem{Guillot02}
Guillot B 2002 {\em Journal of Molecular Liquids\/} {\bf 101} 219 -- 260

\bibitem{Lamoureux-LCP03}
Lamoureux G, Alexander D~MacKerell J and Roux B 2003 {\em The Journal of
  Chemical Physics\/} {\bf 119} 5185--5197

\bibitem{Mankoo-JCP08}
Mankoo P~K and Keyes T 2008 {\em The Journal of Chemical Physics\/} {\bf 129}
  034504

\bibitem{Piquemal-JPC07}
Piquemal J~P, Chelli R, Procacci P and Gresh N 2007 {\em The Journal of
  Physical Chemistry A\/} {\bf 111} 8170--8176

\bibitem{FS2002}
Franzese G and Stanley H~E 2002 {\em Journal of Physics-Condensed Matter\/}
  {\bf 14} 2201--2209

\bibitem{FS-PhysA2002}
Franzese G and Stanley H~E 2002 {\em Physica A-Statistical Mechanics And Its
  Applications\/} {\bf 314} 508--513

\bibitem{FMS2003}
Franzese G, Marques M~I and Stanley H~E 2003 {\em Physical Review E\/} {\bf 67}
  011103

\bibitem{Teixeira1990}
Teixeira J and Bellissent-Funel M~C 1990 {\em Journal of Physics: Condensed
  Matter\/} {\bf 2}

\bibitem{Luzar-Chandler96}
Luzar A and Chandler D 1996 {\em Phys. Rev. Lett.\/} {\bf 76} 928--931

\bibitem{Henry2002}
Henry M 2002 {\em Chemphyschem\/} {\bf 3} 561--9

\bibitem{Baranyai05}
Baranyai A, Bart{\'o}k A and Chialvo A~A 2005 {\em J Chem Phys\/} {\bf 123}
  054502

\bibitem{Smith2004}
Smith J~D, Cappa C~D, Wilson K~R, Messer B~M, Cohen R~C and Saykally R~J 2004
  {\em Science\/} {\bf 306} 851--3

\bibitem{Suresh00}
Suresh S~J and Naik V~M 2000 {\em The Journal of Chemical Physics\/} {\bf 113}
  9727--9732

\bibitem{De03}
Debenedetti P~G 2003 {\em Journal of Physics: Condensed Matter\/} {\bf 15}
  R1669--R1726

\bibitem{Narten67}
Narten A, Danford M and Levy H {1967} {\em {Discussions of the Faraday
  Society}\/}  {97--\&}

\bibitem{Soper1986}
Soper A~K and Phillips M~G 1986 {\em Chemical Physics\/} {\bf 107} 47 -- 60

\bibitem{Odessa2009}
Stokely K, Mazza M~G, Stanley H~E and Franzese G 2010 {\em Metastable Systems
  under Pressure\/} (Springer) chap Metastable Water Under Pressure, pp
  197--216 NATO Science for Peace and Security Series A: Chemistry and Biology

\bibitem{FYS2000}
Franzese G, Yamada M and Stanley H~E 2000 {\em Aip Conference Proceedings\/}
  vol 519 pp 281--287

\bibitem{Franzese:2005sf}
Franzese G and Stanley H~E 2005 {\em Science And Culture Series: Physics\/}
  vol~26 pp 210--214

\bibitem{KumarFS2008}
Kumar P, Franzese G and Stanley H~E 2008 {\em Journal of Physics: Condensed
  Matter\/} {\bf 20} 244114

\bibitem{KFS2008}
Kumar P, Franzese G and Stanley H~E 2008 {\em Physical Review Letters\/} {\bf
  100} 105701

\bibitem{KumarFBS2008}
Kumar P, Franzese G, Buldyrev S~V and Stanley H~E 2008 {\em Aspects of Physical
  Biology\/}  3--22

\bibitem{FranzeseSCKMCS2008}
Franzese G, Stokely K, Chu X~Q, Kumar P, Mazza M~G, Chen S~H and Stanley H~E
  2008 {\em Journal of Physics-Condensed Matter\/} {\bf 20} 494210

\bibitem{Stokely2010}
Stokely K, Mazza M~G, Stanley H~E and Franzese G 2010 {\em Proc Natl Acad Sci U
  S A\/}

\bibitem{Chu:2009hl}
Chu X~q, Faraone A, Kim C, Fratini E, Baglioni P, Leao J~B and Chen S~H 2009
  {\em The Journal of Physical Chemistry B\/} {\bf 113} 5001--5006

\bibitem{MSSSF2009}
Mazza M~G, Stokely K, Strekalova E~G, Stanley H~E and Franzese G 2009 {\em
  Computer Physics Communications\/} {\bf 180} 497--502

\bibitem{Cataudella:1996ko}
Cataudella V, Franzese G, Nicodemi M, Scala A and Coniglio A 1996 {\em Physical
  Review E\/} {\bf 54} 175--189

\bibitem{FDLS2009}
Franzese G and de~los Santos F 2009 {\em J. Phys.: Condens. Matter\/} {\bf 21}
  504107

\bibitem{Kumar:2006zi}
Kumar P, Franzese G, Buldyrev S~V and Stanley H~E 2006 {\em Physical Review
  E\/} {\bf 73} 041505

\bibitem{Koga00}
Koga K, Tanaka H and Zeng X~C 2000 {\em Nature\/} {\bf 2000} 564--567

\bibitem{FC1999}
Franzese G and Coniglio A 1999 {\em Physical Review E\/} {\bf 59} 6409--6412

\bibitem{dlSF2009}
de~los Santos F and Franzese G 2009 {\em Modeling And Simulation Of New
  Materials: Proceedings Of Modeling And Simulation Of New Materials: Tenth
  Granada Lectures\/} ({\em AIP Conf. Proc.\/} vol 1091) ed Marro J, Garrido
  P~L and Hurtado P~I (Granada (Spain): AIP) pp 185--197

\bibitem{truskett:2401}
Truskett T~M, Debenedetti P~G and Torquato S 2001 {\em The Journal of Chemical
  Physics\/} {\bf 114} 2401--2418

\bibitem{KBSGS2005}
Kumar P, Buldyrev S~V, Starr F~W, Giovambattista N and Stanley H~E 2005 {\em
  Physical Review E (Statistical, Nonlinear, and Soft Matter Physics)\/} {\bf
  72} 051503

\bibitem{Elena2010}
Strekalova E~G, Mazza M~G, Stanley H~E and Franzese G {\em in preparation\/}

\bibitem{Ball08}
Ball P {2008} {\em {Chemical Reviews}\/} {\bf {108}} {74--108}

\bibitem{PhysRevE.62.8449}
De~Los~Rios P and Caldarelli G 2000 {\em Phys. Rev. E\/} {\bf 62} 8449--8452

\bibitem{Levy}
Levy Y and Onuchic J~N 2006 {\em Annu. Rev. Biophys. Biomol. Struct.\/} {\bf
  35} 389--415

\bibitem{Dias:2008sp}
Dias C~L, Ala-Nissila T, Karttunen M, Vattulainen I and Grant M 2008 {\em
  Physical Review Letters\/} {\bf 100} 118101--4

\end{thebibliography}

\providecommand{\newblock}{}

\end{document}